# Enhancing Cell Characterization via Hydrodynamic Compression in Suspended Microchannel Resonators

Alberto Martín-Pérez & Daniel Ramos*

Optomechanics Lab, Instituto de Ciencia de Materiales de Madrid (ICMM), CSIC, 3, Sor Juana Inés de la Cruz, 28049 Madrid, Spain

(Dated: January 10, 2025)

**Microfluidics offer remarkable flexibility for in-flow analyte characterization and can even measure the mechanical properties of biological cells through the application of hydrodynamic forces. In this work, we present a new approach to enhance the performance of nanomechanical resonators featuring integrated microfluidic channels when they are used as cell sensors by means of applying hydrostatic compressions. For this purpose, we have studied analytically how this kind of compressions affects either the mechanical properties of the resonator as well as the analytes. We found that, depending on factors such as device geometry and material composition, the mass limit of detection of the resonator can be reduced while the buoyant mass of the particles is increased when a hydrostatic compression is applied, improving the performance of the sensor. Furthermore, we demonstrate that applying these hydrostatic compressions induces shifts in mass distributions among cell lines with similar physical properties, which not only potentially enhances the ability to differentiate between these lines, but also opens the door to measure the cell's compressibility, a biophysical parameter of interest with practical diagnostic applications.**

## Introduction

Early diagnosis is crucial in healthcare, as it enables timely intervention and treatment of specific diseases, avoiding complications and, therefore, increasing the survival rate of the patients and their overall quality of life[1-3]. This is especially important in the context of cancer[4,5]. Early identification of cancer often means smaller tumors and limited metastases, making it more susceptible to existing treatments such as surgery, radiation and chemotherapy[6]. However, developing precise and sensitive diagnostic tools for detecting cancer at its earliest stages is fraught with complexities[7]. The challenges arise from factors such as the heterogeneity of cancer, the small size and localized nature of early-stage tumors or intricate biomarker identification. Despite these hurdles, ongoing research in fields such as genomics[8-11], imaging[12], and liquid biopsy techniques[13-15] is gradually enhancing our capacity to detect cancer in its nascent stages. In this sense, nanotechnology emerges as a very promising technique in advancing the development of innovative diagnostic tools for various diseases[16-18], offering novel ways to detect and study not only biological markers but also physical properties at cellular level. This can be exemplified by the use of plasmonic nanostructures as gold nanoparticles engineered to enhance the sensitivity of sensing devices or to selectively bind to cancer markers[19-21]. Nanoscale biosensors are capable of real-time monitoring of biomarkers[22]; liquid biopsies leveraging nanotechnology can analyze circulating tumor cells, cell-free DNA, and exosomes in the bloodstream, offering a less invasive and more frequent means of cancer detection and monitoring[23].

Among various nanotechnology techniques, nanomechanical sensing stands out as one of the most promising in terms of both sensitivity and versatility. This kind of sensors allows to measure the mechanical properties of biological entities, which has been recognized as very important parameter of tumoral cells as it offers valuable insights into disease progression and potential therapeutic strategies. These properties, such as cell stiffness and deformability, can serve as crucial biomarkers, allowing for the differentiation between cancerous and healthy cells[24]. Tumor cells often exhibit altered mechanical properties compared to their normal counterparts due to changes in their cytoskeleton, membrane composition, and overall architecture[25]. This knowledge can be harnessed for the development of innovative diagnostic techniques, including microfluidic devices and atomic force microscopy, which can detect and characterize cancer cells based on their mechanical signatures[15,26].

In this study, we present a novel approach aimed at enhancing the sensitivity for discriminating between tumoral and healthy cells through the utilization of their varying stiffness characteristics, employing suspended microchannel resonator (SMR) devices. SMR devices integrate nanomechanical resonators with microfluidic channels[27], offering the capability to analyze particles by combining the high sensitivity of nanomechanical resonators[28-31] with the high throughput of microfluidics[15,32], as well as the ability to maintain samples in a liquid environment, which is essential for biological analysis. To assess cell stiffness, we apply hydrostatic

*Correspondence email address: daniel.ramos@csic.es

compressions, a method that does not disrupt the fluid flow. Applying hydrostatic compression to SMR devices has two primary effects. Firstly, it amplifies the differences in certain physical properties (such as size and buoyant mass)

These mass sensors have limitations regarding the minimum detectable mass, as illustrated in Fig. 1a and 1b. This limitation arises due to intrinsic noise present in the of various cell populations being measured. Secondly, it alters the mechanical properties of the resonator. In this context, we investigate the fundamental principles that dictate how this pressure impacts the mechanical properties of the resonator. We then use these findings to determine the limit of detection for this technique and project how its sensitivity might improve for different types of analytes.

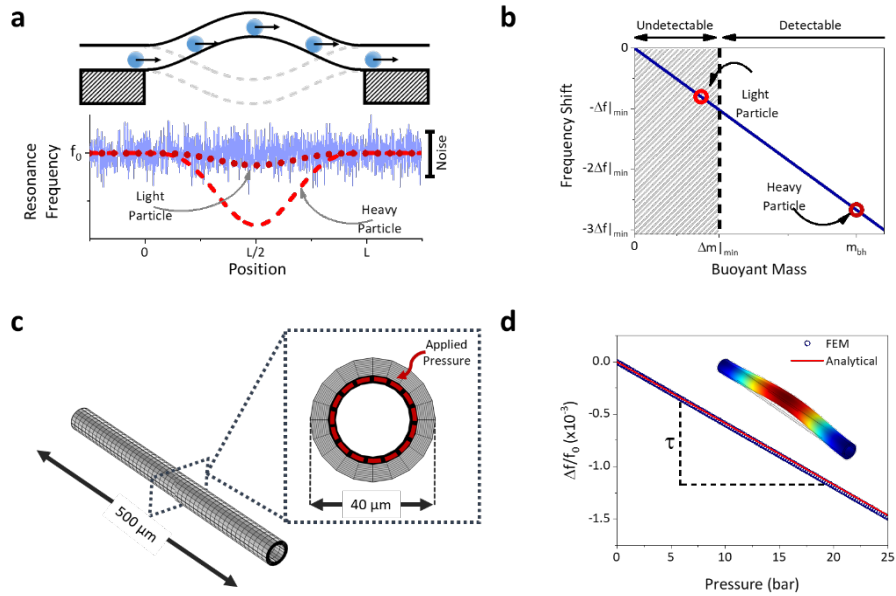

**Figure 1. a.** (Top) Schematic illustrating a particle crossing the suspended microchannel resonator. (Bottom) Resonance frequency plotted against the particle's position on the resonator axis for two particles: one heavy (red dashed line) and the other light (red dotted line). These frequency shifts are compared to the intrinsic frequency noise of the device (blue solid line). **b.** Maximum resonance frequency shift depicted as a function of the buoyant mass of the particle (solid blue line). **c.** Image displaying the geometry and meshing of the resonator simulated via the Finite Element Method (FEM). **d.** Resonance frequency calculated as a function of the applied pressure, obtained through FEM simulations (blue circles) and the analytical model (solid red line). (Inset) Mode shape acquired during the simulation.

## Results
### *Effect of the hydrostatic pressure on mass sensing*
The working principle of SMR devices when employed as mass sensors relies on monitoring their mechanical resonance frequency over time. A deviation in this frequency is recorded as the particle traverses the suspended region (Fig. 1a). This frequency shift is governed by both the buoyant mass of the particle ($m_b$, representing the difference in mass between the particle and the evacuated fluid) and the particle's position along the resonator axis ($x_0$, as expressed in Eq. 1)[33]

$$\frac{\Delta f_n}{f_{0,n}} \approx -\frac{1}{2}\frac{m_b}{m_r}\psi^2(x_0) \qquad (1)$$

Being $\Delta f_n$ the frequency difference in the nth resonance mode, $f_{0,n}$ the resonance frequency of the nth mode of the unloaded resonator, $m_r$ the mass of the resonator and $\psi$ the oscillation mode shape[34].

measured frequency signal. Consequently, the Mass Limit of Detection (MLD) is determined by the minimum detectable frequency shift ($\Delta f/f_0|_{min}$), which, as indicated by previous studies[35], is inversely proportional to the product of twice the quality factor ($Q_n$) and the signal-to-noise ratio (SNR). Considering that the quality factor depends on both the mass and frequency of the resonator ($Q_n = 2\pi m_r f_n/\gamma$, with $\gamma$ representing the dissipation factor), its substitution into Equation 1 yields the following expression for the Mass Limit of Detection ($\Delta m_{min}$).

$$\Delta m_{min} = \frac{\gamma}{2\pi f_n \psi^2(x_0) SNR} \qquad (2)$$

Hence, the determination of this mass limit of detection depends on the experimental conditions, including sources of dissipation and the quality of the signal through the signal-to-noise ratio, in addition to the mechanical properties governing the resonance frequency of the device. The standard method employed in nanomechanical resonators to enhance the MLD involves the simultaneous raising of the resonance frequency and the reduction of the resonator's mass by downscaling the device; however, this approach proves less viable for SMR given the detrimental effect of downscaling the inner channel on flow rates. Nevertheless, previous studies on SMR have demonstrated that both its mass[36] and resonance frequency[37] can be modulated by applying hydrostatic compressions within the inner fluid, introducing a new approach to reduce the MLD. It is noteworthy that hydrostatic compression, characterized by a zero-pressure difference between the inlet and outlet of the channel, can be applied to the inner fluid without affecting flow rates. This ensures that essential parameters for the optimal performance of the device like throughput and particle velocity are maintained[38,39].

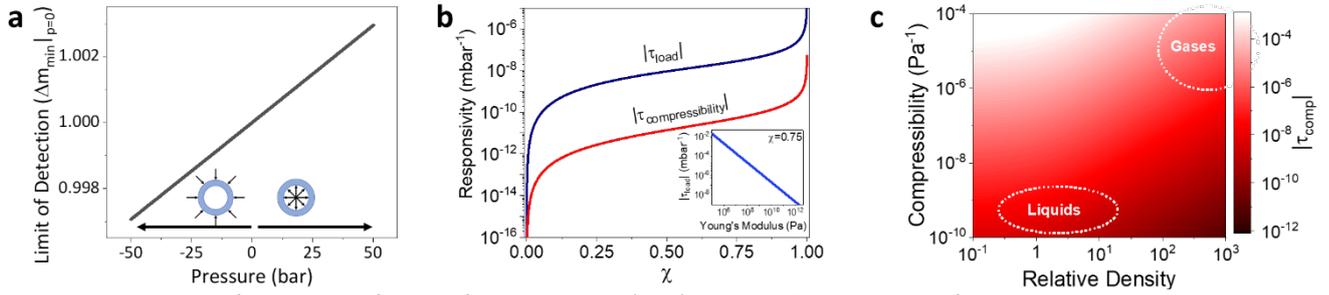

Figure 2: **a.** Mass limit of detection as a function of applied pressure. (Inset) Schematic illustrating the net force acting on the resonator based on the pressure's sign. **b.** Absolute value of the responsivity for the load and compressibility terms (note that both functions have negative sign). (Inset) Absolute value of the load component in the responsivity as a function of the Young's Modulus of the resonator for a constant value of the diameter (0.75). **c.** Absolute value of the compressibility component of the responsivity as a function of the compressibility and density of the fluid. The area marked as "liquids" shows in an illustrative manner typical values of compressibility and density for water and its binary solutions at room temperature and pressure while the area marked as "gases" does the same for ideal gases with molecular masses from 2 to 100 dalton.

The resonance frequency of the SMR as a function of the pressure (Eq. 3) can be calculated by solving the Euler-Bernouilli model[33,40], see Materials and methods for further details.

$$f_n(p) = \frac{\alpha_n^2}{2\pi L^2} \sqrt{\frac{L}{m_0 + V_{in}\rho_f \beta p} \left( EI(p) - \frac{AL^2}{\alpha_n^2} p \right)} \quad (3)$$

Being $E$ the Young's modulus, $I$ the second moment of area, $p$ the applied pressure, $A$ the area of the cross section of the inner channel, $L$ the length of the resonator, $m_0$ the mass of the resonator at null pressure, $V_{in}$ the volume of the inner channel, $\rho_f$ the density of the fluid and $\beta$ the compressibility of the fluid. Note that this expression is equally valid either for singly-clamped or doubly-clamped resonators, which only differs in the value of the modal number ($\alpha_n$).

To validate the model, we conducted simulations using the finite element method (FEM) with commercial software (COMSOL Multiphysics) to determine the resonance frequency of the first flexural mode in a doubly-clamped suspended microchannel resonator. The resonator had an annular cross-section made of fused silica ($E$=73 GPa, $\rho_w$=2200 kg/m³). For this purpose, we modeled hydrostatic pressure as a constant, uniformly distributed pressure along the inner surface of the channel's wall (see Fig. 1c). Additionally, we introduced the mass of the inner fluid as an added mass distributed uniformly along the resonator's axis. Its density varied with the hydrostatic pressure as described in Materials and Methods, using parameters for water ($\rho_f$=998 kg/m³ and $\beta$=0.4 GPa⁻¹).

When comparing the results obtained from the FEM simulations to those of the analytical model (see Fig. 1d), not only do we observe a favorable agreement between the analytical model and the simulations, but it also becomes evident that the resonance frequency can be linearly approximated for small pressures ($p \ll E$). Accordingly, we introduce the responsivity parameter ($\tau$) as the slope of this linear relationship. This parameter offers a more straightforward means to characterize the responsiveness of the resonance frequency to variations in pressure.

$$f_n(p) \approx f_n|_{p=0} \left[ 1 + \frac{1}{2} \left( \frac{1}{I|_{p=0}} \frac{dI|_{p=0}}{dp} - \frac{AL^2}{\alpha_n^2 EI|_{p=0}} - \frac{V_{in}\rho_f \beta}{m_0} \right) p \right] \quad (4)$$

Being $f_n|_{p=0}$ and $I|_{p=0}$ respectively the resonance frequency of the nth mode and the second moment of area at null pressure.

Please note that the responsivity is formed by the sum of three terms, which correspond with the three pressure-dependent parameters introduced in the Euler-Bernouilli equation. These terms are (from left to right in Eq. 4) the moment of area, the hydrostatic load and the compressibility terms. Each of these terms can be studied as a partial responsivity ($\tau_m, \tau_l, \tau_c$ respectively) so the sum of all them gives the total responsivity ($\tau$). Eventually, we obtain an analytical expression for the MLD as a function of the applied pressure by substituting eq. 4 in eq 2.

$$\Delta m_{min}(p) = \frac{\Delta m_{min}|_{p=0}}{1+\tau p} \quad (5)$$

Being $\Delta m_{min}|_{p=0}$ the mass limit of detection at null pressure.

### *Responsivity optimization*
The responsivity's sign significantly impacts MLD, reducing it with positive responsivity under positive pressure and increasing it under negative pressure (Fig. 2a); the opposite holds for negative responsivity. Notably, this model accommodates negative pressures since we have designated external pressure as null (typically atmospheric pressure). Thus, we investigate how responsivity varies with SMR device parameters to optimize MLD reduction. Exploiting the aforementioned fact that responsivity is the linear combination of three parameters due to distinct phenomena, we examine each partial responsivity separately. Equation 4 reveals that the hydrostatic load and compressibility terms are consistently negative, while the inertia term must be positive. To discern how each device parameter influences responsivity's sign, we explore the specific case of an annular cross-section doubly-clamped resonator, obtaining the subsequent values for the load and compressibility terms.

$$\tau_l = -\frac{2\chi^2}{(1-\chi^4)E}\left[\frac{\varphi}{\alpha_n}\right]^2 \quad (6)$$

$$\tau_c = -\frac{\chi^2}{2(\chi^2[1-\rho]+\rho)}\beta \quad (7)$$

Being $\chi$ the relative thickness (i.e., the ratio between the inner and the outer diameters), $\varphi$ the slenderness ratio (i.e., the ratio between the length and the outer diameter) and $\rho$ the relative density (i.e., the quotient between the densities of the resonator's wall [$\rho_w$] and the inner fluid [$\rho_f$]).

Regarding load responsivity, it tends to approach zero rapidly for higher modes and increases in magnitude for slimmer resonators ($\varphi \to \infty$). When graphed against diameter and Young's modulus (Fig. 2b), it is evident that the magnitude of this parameter reaches its peak for thin walls ($\chi \to 1$) and soft materials ($E \to 0$). Additionally, the compressibility component in the responsivity directly follows a behavior similar to that of the diameter (Fig. 2b). The difference is that for thin walls, its value tends to -β/2. Moreover, this term is directly proportional to the fluid's properties. When plotting the value of this partial responsivity for a fixed diameter as a function of the fluid's density and compressibility (Fig. 2c), it becomes evident that its value is maximized for higher compressibilities and densities. Consequently, under conditions of room temperature and atmospheric pressure, this term is minimized for liquids and maximized for gases.

In the case of the inertia component, obtaining an analytical expression is complex due to the inverse proportionality of volume changes to the bulk modulus ($K$) of the resonator wall. This complexity arises from the challenge of pinpointing the contribution of each radius's expansion to this volume variation. To address this, we introduce expansion functions for the inner and outer tube diameters ($\varepsilon_i(\chi)$ and $\varepsilon_o(\chi)$, a formal definition of these functions can be found at Materials and Methods), allowing us to derive the second moment of area analytically (Eq. 8) and, consequently, an expression for the inertia component of the responsivity (Eq. 9).

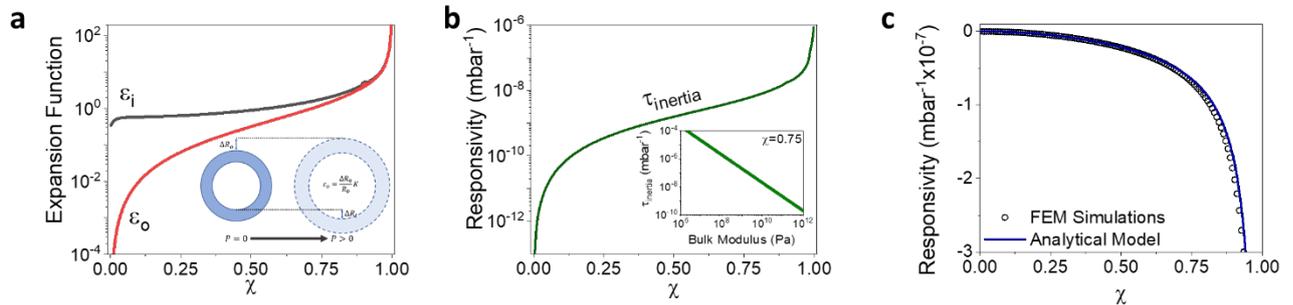

**Figure 3. a**. Results derived from FEM simulations for the expansion functions. (Inset) Schematic demonstrating the method for measuring these functions. **b**. Inertia term of the responsivity as a function the diameter. (Inset) Inertia term of the responsivity as a function of bulk modulus for a fixed diameter (0.75). **c.** Total responsivity of the device as a function of relative thickness, obtained from both FEM simulations (circles) and the analytical model (solid line).

$$I(p) = \frac{\pi}{4}R_0^4\left(\left[1-\varepsilon_o(\chi)\frac{p}{K}\right]^4 - \chi^4\left[1-\varepsilon_i(\chi)\frac{p}{K}\right]^4\right) \quad (8)$$

$$\tau_{inertia} = \frac{2[\varepsilon_o(\chi)-\chi^4\varepsilon_i(\chi)]}{(1-\chi^4)K} \quad (9)$$

To find these expansion functions, we conducted additional FEM simulations to assess how the inner and outer radii change in response to applied pressure and thickness. (Fig. 3a, see Supplementary information for further details on these simulations). When plotting the inertia component as a function of the thickness and the bulk modulus of the resonator, Fig. 3b, it becomes evident that it approaches zero rapidly when the relative thickness deviates from 1 and as the bulk modulus increases. Please note that the bulk modulus is directly proportional to Young's modulus, $K = E/(3[1-2\nu])$, being ν the Poisson ratio.

To validate the theoretical model, we compare it with values obtained through FEM simulations. For a given thickness, we determine the eigenfrequency as a function of applied pressure and calculate the total responsivity by fitting a linear function. This process is repeated for various thicknesses, allowing us to establish the relationship between responsivity and thickness. The responsivity results from FEM simulations exhibit strong agreement with those obtained using the analytical model, Fig. 3c.

Additionally, upon comparing the analytical value of the load term's partial responsivity with the FEM results, the ratio of 0.4999[42]). Consequently, substantial variations in buoyant mass occur when they are compressed. This load term is found to dominate across all thickness values, while the inertia and compressibility terms act as second-order effects.

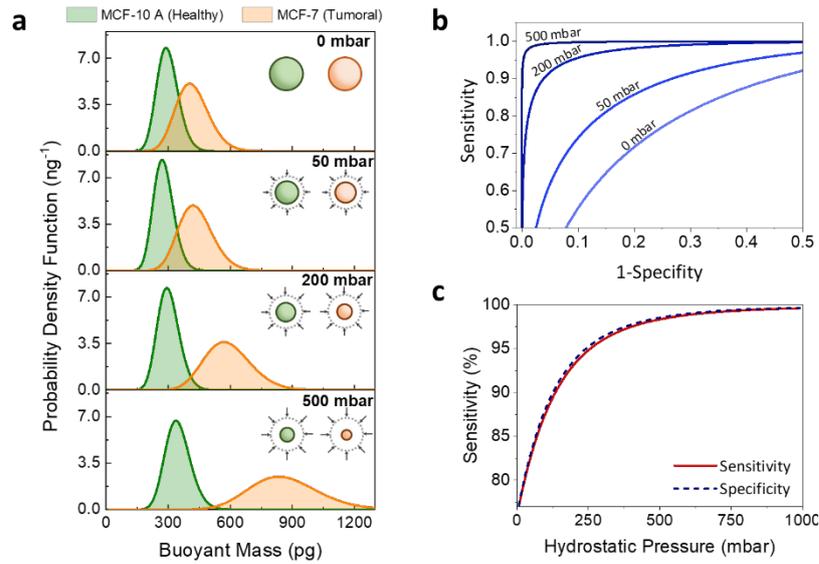

**Figure 4. a.** Mass distributions calculated for the cell lines. Inset. Schematic of the size variation of the cells as a function of the applied pressure (not to scale) **b.** ROC curves for some different applied pressures. Please, note that for the sake of clarity just 4 curves of a total of 100 are shown in this plot. **c.** Sensitivity as a function of the applied pressure.

*Effect of the hydrostatic pressure on buoyant mass*
Hydrostatic compressions not only affect the mechanical properties of the SMR device but also impact the particles under analysis, causing a reduction in their volume, which, in turn, affects the sensor's performance. In a linear regime, when hydrostatic pressure is applied, the volume of the particles varies as $\Delta V_p/V_{p0} = -p/K_p$, where $\Delta V_p$ represents the change in particle volume, $V_{p0}$ is the volume of the particle at null pressure, and $K_p$ denotes the particle's bulk modulus (Fig. 4a, insets). During these hydrostatic compressions, the particle's mass remains constant, while the volume of the evacuated liquid decreases, consequently, the buoyant mass of the particle is changed as

$$m_b(p) = m_b|_{p=0}\left(1 + \frac{\rho_f}{[\rho_p-\rho_f]K_p}p\right) \quad (10)$$

Being $m_b|_{p=0}$ the buoyant mas of the particle at null pressure and $\rho_p$ the density of the particle.

This variation in buoyant mass can be disregarded for particles made of most materials since their bulk moduli ($K_p$) are sufficiently high ($K_p > 1\ GPa$) to not induce significant changes in particle volume within the typical pressure range used in these devices[36] (<1 MPa). However, this does not hold true for biological cells[41], whose bulk moduli range from ~0.1 MPa to ~100 MPa (assuming Young's moduli from ~0.1 kPa to ~100 kPa[42] and a Poisson pressure-assisted modification of the buoyant mass opens the door to enhance the cell discerning in these devices because in many cases, although the precision is sufficiently high for mass sensing of a specific type of cells, distinguishing between two different types of cells becomes impossible if their buoyant mass distributions overlap. This issue arises from the inherent variability in size among cells of the same type, resulting in a broad buoyant mass distribution. This distribution depends on the cell's mass density and size distribution as follows[15]

$$PDF(m_b) = \frac{\mu}{6\sigma\sqrt{\pi}}m_b^{-2/3}e^{-\left(\mu\frac{\sqrt[3]{m_b}-\sqrt[3]{m_{av}(p,K_p)}}{2\sigma}\right)^2} \quad (11)$$

Being $PDF$ the probability density function, $\mu = \left(4\pi[\rho_p-\rho_f]/3\right)^{-1/3}$, $\sigma$ the standard deviation of the size distribution and $m_{av}(p,K_p)$ the buoyant mass of the average-size cell calculated as a function of the applied pressure and cell's bulk modulus as shown in equation 10, note that this function is normalized so its integral is 1.

The overlapping of buoyant mass distributions presents a classic challenge in SMR devices when distinguishing between two different cell lines from the same tissue, such as a healthy line and a cancerous line[15]. Due to the significant similarities in size and mass density between these lines, there is nearly complete overlap of the distributions (Fig. 4a at 0 mbar), rendering both lines indistinguishable. This limitation has hindered the general use of SMR devices for cellular diagnosis. However, the influence of hydrostatic compressions offers a new mean to differentiate cells, even in such cases, as previous studies have indicated variations in stiffness among different cell lines[24]. Consequently, hydrostatic compressions can be used to adjust the extent of overlap between the distributions, as one of the two cell lines will respond more readily to changes in pressure.

To quantify the improved capacity for differentiation with applied pressure, we calculate the buoyant mass distribution, Fig. 4a, for two cell lines from the same tissue, specifically using data from the MCF-10A (healthy) and MCF-7 (tumoral) lines, as they have been extensively studied in previous literature and specifically in works involving SMR devices[15,36]. Please, refer to table 1 for the relevant data and their references. Having computed the mass distributions as a function of applied pressure, we proceed to calculate the receiver operating characteristic (ROC) curve for each pair of mass distributions at various applied pressures, Fig. 4b (see Materials and Methods for further details in the calculation of these curves). Eventually, we determine the optimal sensitivity and specificity for each ROC curve as the point closest to 100% sensitivity and specificity. This allows us to observe the trend as a function of applied pressure (Fig. 4c). Following this analysis, it becomes evident that the application of hydrostatic compressions enhances the discriminatory capacity of the SMR sensors, tending toward unambiguous discrimination (i.e., achieving 100% sensitivity and specificity) at high pressures.

|  | $\rho_p$ (g/mL) | $\sigma$ (μm) | Average radius (μm) | Young's Modulus (MPa) | Poisson Ratio |
|---|---|---|---|---|---|
| MCF-10A | 1.1 | 2 | 17 | 1 | 0.4999 |
| MCF-7 | 1.1 | 2 | 19 | 0.1 | 0.4999 |
| Reference | 39 | 15 | 15 | 43 | 43 |

**Table 1.** Physical parameters used for calculate the buoyant mass distributions of the cell lines. Bulk modulus is calculated as $K_p = E/[3(1-2\nu)]$, being ν the Poisson Ratio.

Nevertheless, this optimization is not the unique advantage of applying hydrostatic compressions in SMR devices, it also opens the door to measure cell stiffness if the mass distribution of the same sample is measured at different hydrostatic compressions. In that case, the bulk modulus of the cells can be obtained from the slope of the buoyant mass as a function of the applied hydrostatic pressure (Eq. 10).

## Discussion

In this study, we analytically examined the impact of hydrostatic pressure within a nanomechanical resonator with an integrated channel on various sensing parameters, including the mass limit of detection, sensitivity, and specificity. We found that these parameters could be finely tuned by adjusting the device's responsivity (the slope of the resonance frequency-pressure relationship) and the bulk modulus of the particles under analysis. Responsivity, influenced by factors like the device's geometry and material composition, significantly affects the mass limit of detection. Positive responsivity or the application of negative pressure can reduce the mass limit of detection. The magnitude and sign of responsivity depend on the SMR device's properties. Finally, we observed that these compressions can also shift mass distributions among cell lines with similar physical properties, not only enhancing the capability to distinguish between these lines but also allowing to obtain cell stiffness. This effect holds potential for practical diagnostic applications.

## Acknowledgements

This study was partially supported by the Spanish Ministry of Science under projects OMNIPATH (reference PID2020-119587RB-I00) and Q-BOMSAI (reference PDC2023-145919-I00) and partially supported by the European Commission – NextGenerationEU (Regulation EU 2020/2094), through CSIC's Quantum Technologies Platform (QTEP)

## Materials and methods

### Development of the expression for the resonance frequency

For this purpose, we use the Euler-Bernouilli model, taking into consideration the force applied by the hydrostatic pressure in the resonator, which allows us to obtain the following differential equation.

$$EI(p)\frac{d^4w}{dx^4} - pA\frac{d^2w}{dx^2} + \frac{m_r(p)}{L}\frac{d^2w}{dt^2} = 0 \quad \text{(M.1)}$$

Being $E$ the Young's modulus, $I$ the second moment of area, $p$ the applied pressure, $w$ the strain of the resonator, $A$ the area of the cross section of the inner channel, $m_r$ the mass of the resonator and $L$ the length of the resonator.

Please, note that all the three terms in the previous equation depend either implicitly or explicitly on the applied hydrostatic pressure. In the case of the mass of the resonator, the mass variation (Eq. M.2) will be produced by the compressibility of the inner fluid ($\beta$).

$$m_r(p) = m_0 + V_{in}\rho_f(1 + \beta p) \quad \text{(M.2)}$$

Being $m_0$ the mass of the resonator at null pressure, $V_{in}$ the volume of the inner channel and $\rho_f$ the density of the fluid.

On the other hand, the hydrostatic compressions will change either the volume of the inner channel as well as the resonator wall, producing the variation of the second moment of area. However, since the expression of this parameter varies strongly depending on the geometry of the cross section, we will keep it in its implicit form for the sake of the generality of the model.

The previous differential equation (Eq. M.1) can be resolved in an analogous manner to the Euler-Bernouilli model without external forces, obtaining the same mode shape and modal numbers but different resonance frequency (Eq. 5).

*FEM simulations*
To simulate the effect of the pressure in the resonance frequency we calculate the eigenfrequency of the resonator described in Fig. 1.c. for different hydrostatic pressure values. For this simulation the resonator consists only on its solid wall while the part that contains the inner fluid remains empty. To consider the effect of the fluid's mass on the resonance frequency, we included an added mass distributed homogeneously over the surface of the inner face. The value of this added mass was introduced as shown in the second term in equation M.2.

For a given diameter we calculate the responsivity as the slope of the resonance frequency as a function of the applied pressure and we repeat the process for different values of the relative thickness ranging from 0 to 1, obtaining the values shown in Fig. 3c.

*Expansion functions*
For obtaining the value of the expansion functions we have simulated the annular cross section of an SMR device with an outer diameter of 40 μm made of fused silica, including the hydrostatic pressure as a constant and isotropic force over the frontier between the wall and the inner fluid. We calculate the difference between the inner and outer diameters with their respective unloaded values ($\Delta R_i$ and $\Delta R_o$) for different values of the pressure ranging from -1 mbar to 1 mbar. The expansion functions are calculated as shown in eq. M4.

$$\frac{\varepsilon_o}{K} = \frac{\Delta R_o}{R_0}\Delta p \qquad (M.4)$$

We repeat this process iteratively for different values of $\chi$ ranging from 0 to 1, which allows us to obtain the functions plotted in Fig. 2d.

*Calculation of the ROC curves*
For the calculation of the ROC curves we have developed a software in Matlab so we can calculate the mass distributions as shown in equation 11 for the both cell lines at different values of the pressure using the parameters shown in table 1. For each mass distribution pair, we set a mass discrimination criterion ($m_{threshold}$) which is swept from 10 pg to 2000 pg, dividing the curves in their lower part (from 10 pg to $m_{threshold}$) and their upper part (from $m_{threshold}$ to 2000 pg). Consequently, we can calculate the false negative rate (FN) as the integral of the lower part of the MCF-7 mass distribution and the true positive (TP) rate as the integral of the upper part of this mass distribution.

Analogously, we calculate the true negative (TN) rate as the integral of the lower part of the mass distribution of the MCF-10A distribution and the false positive (FP) rate as the integral of the upper part of this function. Eventually, these data are used to obtain the sensitivity (TP/[TP+FN]) and specificity (TN/[TN+FP]) values.

## References

1    Myszczynska, M. A. *et al.* Applications of machine learning to diagnosis and treatment of neurodegenerative diseases. *Nature Reviews Neurology* **16**, 440-456, doi:10.1038/s41582-020-0377-8 (2020).
2    Soong, J. & Soni, N. Sepsis: recognition and treatment. *Clinical Medicine* **12**, 276-280, doi:10.7861/clinmedicine.12-3-276 (2012).
3    Nordberg, A. Towards early diagnosis in Alzheimer disease. *Nature Reviews Neurology* **11**, 69-70, doi:10.1038/nrneurol.2014.257 (2015).
4    Crosby, D. *et al.* Early detection of cancer. *Science* **375**, eaay9040, doi:doi:10.1126/science.aay9040 (2022).
5    Srinivas, P. R., Barker, P. & Srivastava, S. Nanotechnology in Early Detection of Cancer. *Laboratory Investigation* **82**, 657-662, doi:10.1038/labinvest.3780460 (2002).
6    Etzioni, R. *et al.* The case for early detection. *Nature Reviews Cancer* **3**, 243-252, doi:10.1038/nrc1041 (2003).
7    Wang, L. Early Diagnosis of Breast Cancer. *Sensors* **17**, 1572 (2017).
8    Mardis, E. R. The emergence of cancer genomics in diagnosis and precision medicine. *Nature Cancer* **2**, 1263-1264, doi:10.1038/s43018-021-00305-6 (2021).
9    Navin, N. E. Cancer genomics: one cell at a time. *Genome Biology* **15**, 452, doi:10.1186/s13059-014-0452-9 (2014).
10   Kosaka, P. M. *et al.* Detection of cancer biomarkers in serum using a hybrid mechanical and optoplasmonic nanosensor. *Nature Nanotechnology* **9**, 1047-1053, doi:10.1038/NNANO.2014.250 (2014).
11   Mertens, J. *et al.* Label-free detection of DNA hybridization based on hydration-induced tension in nucleic acid films. *Nature Nanotechnology* **3**, 301-307, doi:10.1038/nnano.2008.91 (2008).
12   Frangioni, J. V. New Technologies for Human Cancer Imaging. *Journal of Clinical Oncology* **26**, 4012-4021, doi:10.1200/jco.2007.14.3065 (2008).
13   Yu, D. *et al.* Exosomes as a new frontier of cancer liquid biopsy. *Molecular Cancer* **21**, 56, doi:10.1186/s12943-022-01509-9 (2022).

14  Chen, M. & Zhao, H. Next-generation sequencing in liquid biopsy: cancer screening and early detection. *Human Genomics* **13**, 34, doi:10.1186/s40246-019-0220-8 (2019).

15  Martín-Pérez, A. *et al.* Mechano-Optical Analysis of Single Cells with Transparent Microcapillary Resonators. *ACS Sensors* **4**, 3325-3332, doi:10.1021/acssensors.9b02038 (2019).

16  Jin, C., Wang, K., Oppong-Gyebi, A. & Hu, J. Application of Nanotechnology in Cancer Diagnosis and Therapy - A Mini-Review. *International Journal of Medical Sciences* **17**, 2964-2973, doi:10.7150/ijms.49801 (2020).

17  Pericleous, P. *et al.* Quantum dots hold promise for early cancer imaging and detection. *International Journal of Cancer* **131**, 519-528, doi:https://doi.org/10.1002/ijc.27528 (2012).

18  Dominguez, C. M. *et al.* Hydration Induced Stress on DNA Mono layers Grafted on Microcantilevers. *Langmuir* **30**, 10962-10969, doi:10.1021/la501865h (2014).

19  Choi, H. S. *et al.* Design considerations for tumour-targeted nanoparticles. *Nature Nanotechnology* **5**, 42-47, doi:10.1038/nnano.2009.314 (2010).

20  Kingsley, J. D. *et al.* Nanotechnology: A Focus on Nanoparticles as a Drug Delivery System. *Journal of Neuroimmune Pharmacology* **1**, 340-350, doi:10.1007/s11481-006-9032-4 (2006).

21  Ramos, D., Malvar, O., Davis, Z. J., Tamayo, J. & Calleja, M. Nanomechanical Plasmon Spectroscopy of Single Gold Nanoparticles. *Nano Letters* **18**, 7165-7170, doi:10.1021/acs.nanolett.8b03236 (2018).

22  Haes, A. J., Chang, L., Klein, W. L. & Van Duyne, R. P. Detection of a Biomarker for Alzheimer's Disease from Synthetic and Clinical Samples Using a Nanoscale Optical Biosensor. *Journal of the American Chemical Society* **127**, 2264-2271, doi:10.1021/ja044087q (2005).

23  Li, W. *et al.* Emerging Nanotechnologies for Liquid Biopsy: The Detection of Circulating Tumor Cells and Extracellular Vesicles. *Advanced Materials* **31**, 1805344, doi:https://doi.org/10.1002/adma.201805344 (2019).

24  Plodinec, M. *et al.* The nanomechanical signature of breast cancer. *Nature Nanotechnology* **7**, 757-765, doi:10.1038/nnano.2012.167 (2012).

25  Swaminathan, V. *et al.* Mechanical Stiffness Grades Metastatic Potential in Patient Tumor Cells and in Cancer Cell Lines. *Cancer Research* **71**, 5075-5080, doi:10.1158/0008-5472.Can-11-0247 (2011).

26  Cross, S. E. *et al.* AFM-based analysis of human metastatic cancer cells. *Nanotechnology* **19**, 384003, doi:10.1088/0957-4484/19/38/384003 (2008).

27  Burg, T. P. & Manalis, S. R. Suspended microchannel resonators for biomolecular detection. **83**, 2698-2700, doi:10.1063/1.1611625 (2003).

28  Lee, J., Shen, W., Payer, K., Burg, T. P. & Manalis, S. R. Toward Attogram Mass Measurements in Solution with Suspended Nanochannel Resonators. *Nano Letters* **10**, 2537-2542, doi:10.1021/nl101107u (2010).

29  Martín-Pérez, A. & Ramos, D. Nanomechanical hydrodynamic force sensing using suspended microfluidic channels. *Microsystems & Nanoengineering* **9**, 53, doi:10.1038/s41378-023-00531-1 (2023).

30  Calmo, R. *et al.* Monolithic glass suspended microchannel resonators for enhanced mass sensing of liquids. *Sensors and Actuators B: Chemical* **283**, 298-303, doi:https://doi.org/10.1016/j.snb.2018.12.019 (2019).

31  Pastina, A. D., Maillard, D. & Villanueva, L. G. Fabrication of suspended microchannel resonators with integrated piezoelectric transduction. *Microelectron. Eng.* **192**, 83-87, doi:10.1016/j.mee.2018.02.011 (2018).

32  Stockslager, M. A. *et al.* Rapid and high-precision sizing of single particles using parallel suspended microchannel resonator arrays and deconvolution. *Review of Scientific Instruments* **90**, doi:10.1063/1.5100861 (2019).

33  Ruz, J. J. *et al.* A Review on Theory and Modelling of Nanomechanical Sensors for Biological Applications. *Processes* **9**, 164 (2021).

34  Ramos, D. *et al.* Detection of bacteria based on the thermomechanical noise of a nanomechanical resonator: origin of the response and detection limits. *Nanotechnology* **19**, doi:10.1088/0957-4484/19/03/035503 (2008).

35  Sadeghi, P., Demir, A., Villanueva, L. G., Kähler, H. & Schmid, S. Frequency fluctuations in nanomechanical silicon nitride string resonators. *Physical Review B* **102**, 214106, doi:10.1103/PhysRevB.102.214106 (2020).

36  Martín-Pérez, A., Ramos, D., Tamayo, J. & Calleja, M. Nanomechanical Molecular Mass Sensing Using Suspended Microchannel Resonators. **21**, 3337 (2021).

37  Khan, M. F., Knowles, B., Dennison, C. R., Ghoraishi, M. S. & Thundat, T. Pressure modulated changes in resonance frequency of microchannel string resonators. **105**, 013507, doi:10.1063/1.4889744 (2014).

38  Stockslager, M. A. *et al.* Rapid and high-precision sizing of single particles using parallel suspended microchannel resonator arrays and deconvolution. *Review of Scientific Instruments*

**90**, 085004-085013, doi:10.1063/1.5100861 (2019).

39  Martín-Pérez, A. *et al.* Hydrodynamic assisted multiparametric particle spectrometry. *Scientific Reports* **11**, 3535, doi:10.1038/s41598-021-82708-0 (2021).

40  Salapaka, M. V., Bergh, H. S., Lai, J., Majumdar, A. & McFarland, E. Multi-mode noise analysis of cantilevers for scanning probe microscopy. **81**, 2480-2487, doi:10.1063/1.363955 (1997).

41  Moeendarbary, E. & Harris, A. R. Cell mechanics: principles, practices, and prospects. **6**, 371-388, doi:https://doi.org/10.1002/wsbm.1275 (2014).

42  Geltmeier, A. *et al.* Characterization of Dynamic Behaviour of MCF7 and MCF10A Cells in Ultrasonic Field Using Modal and Harmonic Analyses. *PLOS ONE* **10**, e0134999-e0135018, doi:10.1371/journal.pone.0134999 (2015).

43  Yubero, M. L. *et al.* Effects of energy metabolism on the mechanical properties of breast cancer cells. *Communications Biology* **3**, 590, doi:10.1038/s42003-020-01330-4 (2020).